\documentclass[apjl,iop]{emulateapj}

\usepackage{natbib}
\usepackage{graphicx}
\usepackage{txfonts}
\begin{document}

\title{Bayesian Analysis of Multiple Harmonic Oscillations in the Solar Corona}

\shorttitle{Bayesian Analysis of Multiple Harmonic Oscillations in the Solar Corona}

\shortauthors{Arregui, Asensio Ramos, \& D\'{\i}az }

\author{I. Arregui\altaffilmark{1,2}, A. Asensio Ramos\altaffilmark{1,2}, and A. J. D\'{\i}az\altaffilmark{1,2}}
\altaffiltext{1}{Instituto de Astrof\'{\i}sica de Canarias, V\'{\i}a L\'actea s/n, E-38205 La Laguna, Tenerife, Spain}
\altaffiltext{2}{Departamento de Astrof\'{\i}sica, Universidad de La Laguna, E-38205 La Laguna, Tenerife, Spain}

 \email{iarregui@iac.es}

\begin{abstract}
The detection of multiple mode harmonic kink oscillations in coronal loops enables to obtain information on coronal density stratification and magnetic field expansion using seismology inversion techniques. The inference is based on the measurement of the period ratio between the fundamental mode and the first overtone and theoretical results for the period ratio under the hypotheses of coronal density stratification and magnetic field expansion of the wave guide. We  present a Bayesian analysis of multiple mode harmonic oscillations for the inversion of the density scale height and magnetic flux tube expansion, under each of the hypotheses.  Then, the two models are compared using a Bayesian model comparison scheme to assess how plausible each one is, given our current state of knowledge.
\end{abstract}

\keywords{magnetohydrodynamics (MHD) --- methods: statistical --- Sun: corona --- Sun: oscillations}

\section{Introduction}

Observations of magnetohydrodynamic (MHD) oscillations in solar coronal loops indicate the 
simultaneous presence of multiple mode harmonics in the same structure \citep{verwichte04,demoortel07,vandoorsselaere07c,oshea07,vandoorsselaere09}. 
Their potential use for the diagnostic of the plasma conditions in the corona has attracted 
considerable interest \citep[see][for a review]{andries09b}. The idea was first put forward by 
\cite{andries05b} and \cite{goossens06} who found that the ratio of the fundamental mode period 
to twice that of its first overtone in the longitudinal direction depends on the density structuring 
along magnetic field lines. It is equal to unity in longitudinally uniform and unbounded tubes, but 
is smaller than  one when density stratification is present. Using observational measurements of 
period ratios, \cite{andries05b} and \cite{vandoorsselaere07c} obtained estimates for the coronal 
density scale height.

An alternative hypothesis was formulated by \cite{verth08a} whereby the expansion of the 
magnetic loop  produces a deviation from unity in the period ratio, increasing its value. The effect 
is relevant for obtaining accurate estimates of the density scale height using multiple mode period 
ratios \citep{verth08a,verth08b,ruderman08}. Observational evidence for magnetic tube expansion 
has been reported by e.g., \cite{klimchuk00} in soft X-ray loops observed with Yohkoh and by 
\cite{watko00} in non-flare and postflare loops observed with TRACE. Observations of period ratios 
larger than one have been reported by \cite{oshea07} and \cite{demoortel07}.

Since then, period ratio studies have analyzed different models for the density structuring in coronal loops \citep{mcewan06,mcewan08}, the influence of  the elliptic shape of loops \citep{morton09}, the twist of the magnetic field \citep{karami12}, the effect of the environment \citep{orza12}, and the temporal expansion of loops \citep{ballai12}. Period ratios have also been analyzed for slow MHD modes \citep{macnamara10} and for sausage modes \citep{macnamara11} and their use has  been suggested in the context of prominence seismology \citep{diaz10,arregui12a}.

\begin{figure*}[t]
   \centering
   \includegraphics[angle=0,scale=0.48]{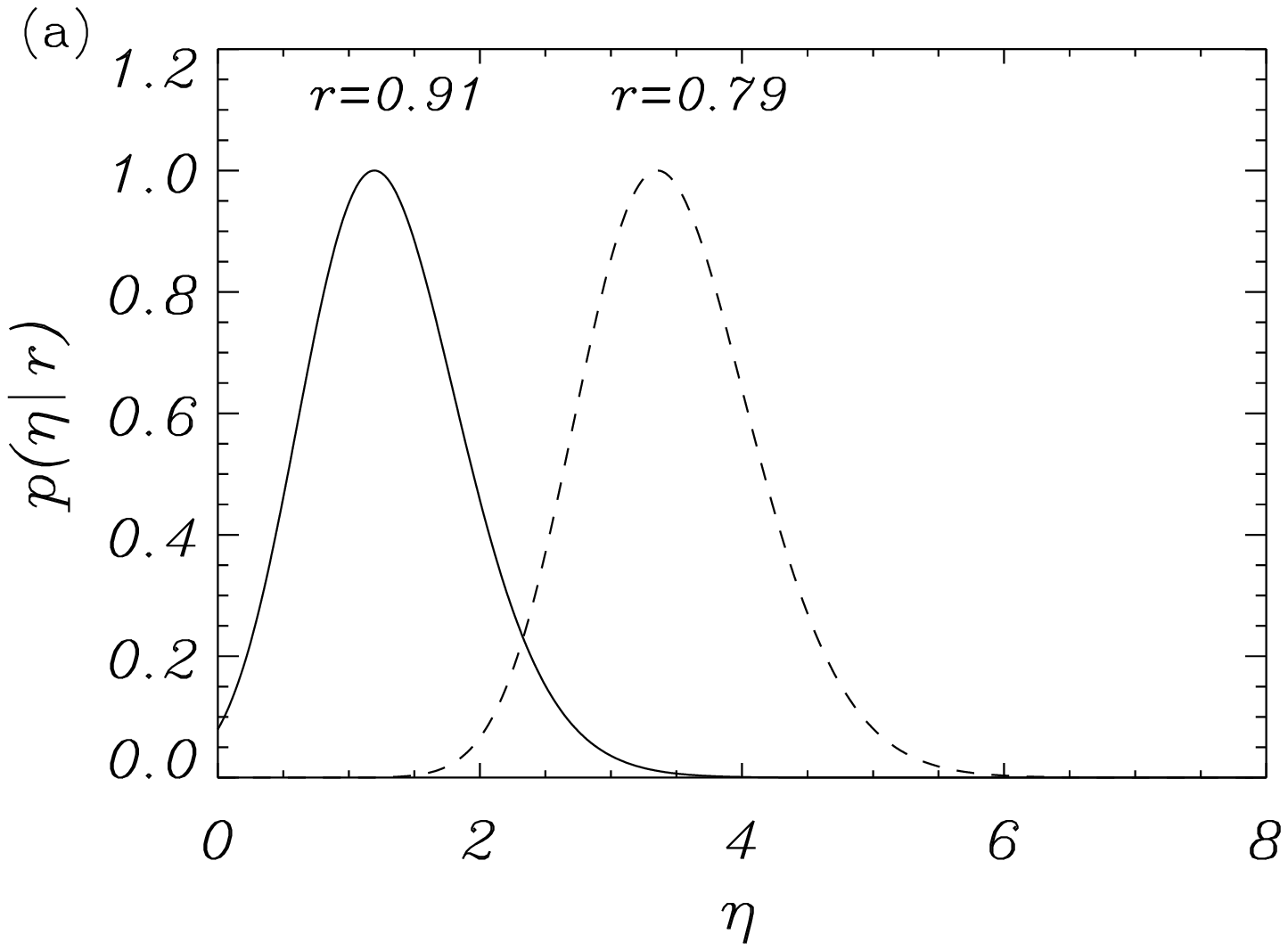} 
     \includegraphics[angle=0,scale=0.48]{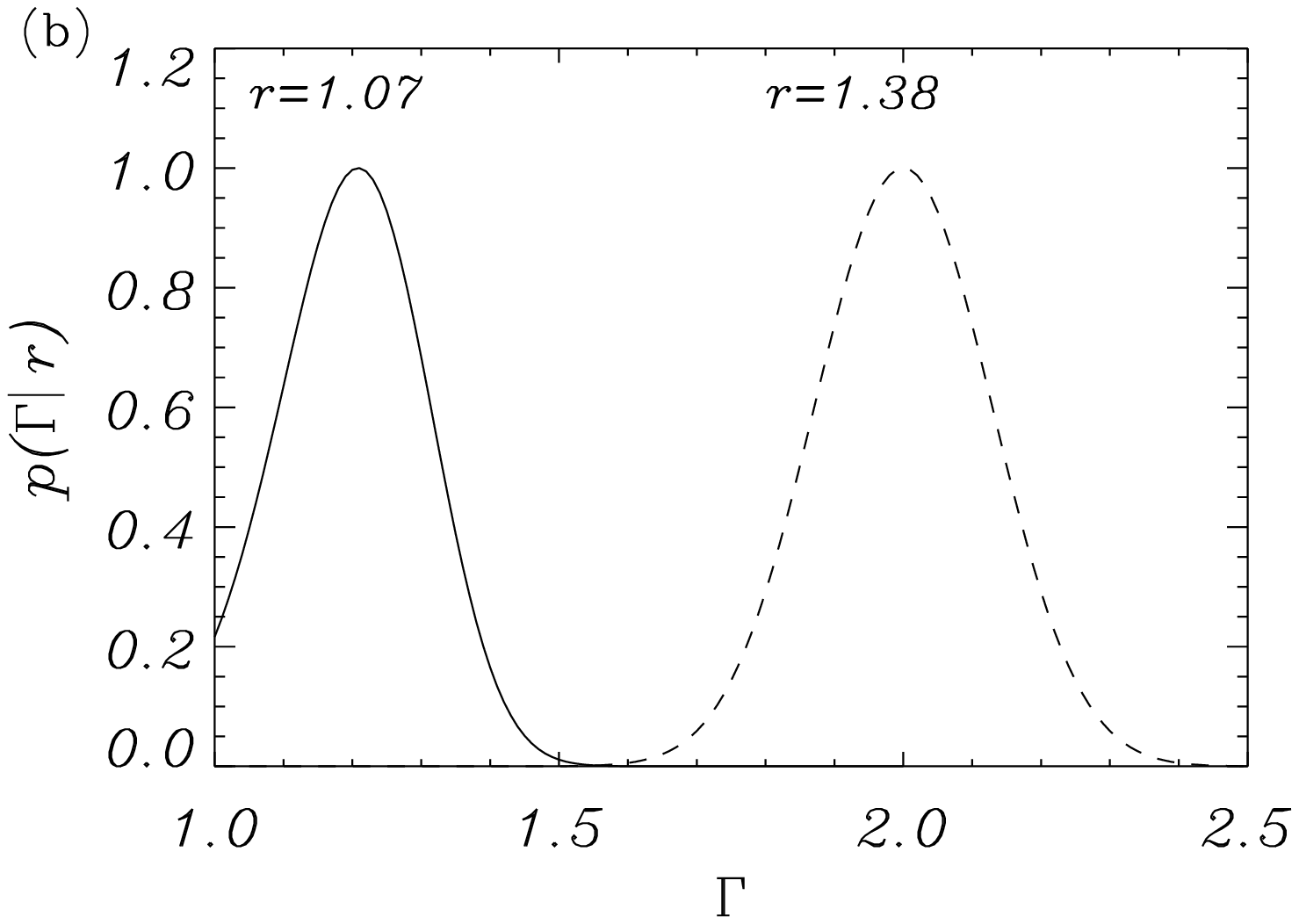} 
  \caption{(a) Posterior distributions for $\eta$, under the density stratification model, for two values of the period ratio. (b) Posterior distributions for the magnetic tube expansion, $\Gamma$, under the magnetic expansion model, for two values of the period ratio. The measured period ratio and the inferred median of the distribution, with uncertainties given at the 68\% credible interval are: (a) $r=0.91\pm0.04$, $\eta=1.26^{+0.65}_{-0.59}$; $r=0.79\pm0.03$, $\eta=3.39^{+0.72}_{-0.64}$. (b)  $r=1.07\pm0.04$, $\Gamma=1.20^{+0.10}_{-0.12}$; $r=1.38\pm0.04$, $\Gamma=1.87^{+0.07}_{-0.07}$. The improved errors on $r$ in the measurements of \cite{verwichte04} have been taken from \cite{vandoorsselaere07c}. Similar errors are assigned to the mean values in the measurements by \cite{demoortel07}. }
    \label{fig:posteriors}
 \end{figure*}

The application of Bayesian analysis techniques to coronal seismology is in its infancy. A first attempt of parameter inference using damped loop oscillations was presented by \cite{arregui11b}.  In this paper, we present the first application of Bayesian model comparison techniques to coronal seismology. We first perform Bayesian parameter inference for the coronal density scale height and the magnetic tube expansion, under the two hypotheses of density stratification and magnetic field divergence, using multiple mode oscillations. Then, we assess which one of the two hypotheses better explains the observations, for given values of the period ratio.

\section{Theoretical Models }

The deviation from unity for the ratio between the fundamental and the first overtone transverse kink oscillation periods has been attributed to two main physical effects.  Each one constitutes a hypothesis to explain the data. 

In the model by \cite{andries05b}  coronal density stratification produces a decrease of the period ratio. This model projects a vertically stratified isothermal atmosphere onto a semicircular loop. An analytical expression for the dependence of the period ratio on density scale height was obtained by \cite{safari07}. This expression can be rewritten as

\begin{equation}\label{forward1}
r_1=\frac{P_1}{2P_2}=1-\frac{4}{5}\left(\frac{\eta}{\eta+3\pi^2}\right),
\end{equation}

\noindent
with $P_1$ and $P_2$ the periods of the fundamental and first overtone modes, $\eta=L/\pi H$ the ratio of the loop height at the apex to the density scale height $H$, and  $L$ the loop length.  We have checked that Equation~(\ref{forward1}) provides us with a good approximation to the numerical results by \cite{andries05b}. 

In the model by \cite{verth08a} magnetic tube expansion produces an increase of the period ratio given by 

\begin{equation}\label{forward2}
r_2=\frac{P_1}{2P_2}=1+\frac{3(\Gamma^2-1)}{2\pi^2},
\end{equation}
with the expansion defined as $\Gamma=r_a/r_f$, where $r_a$ is the radius at the apex and $r_f$ is the radius at the footpoint.  Equation~(\ref{forward2}) was obtained under the assumption that $|\Gamma-1|\ll 1$. Moreover, it is the result for a particular expanding loop model. Another equilibrium state will produce a quantitatively different result, although \cite{ruderman08} anticipate a qualitatively similar result.

Both longitudinal stratification and magnetic tube expansion have forward models that relate one observed quantity, the period ratio $P_1/2P_2$, to one physical quantity to be inferred, $\eta$ or $\Gamma$.

\section{Bayesian Parameter Inference}

 To perform the inference using existing estimates for the period ratio we employ  Bayes' theorem \citep{bayes73}

\begin{equation}
p(\mbox{{\boldmath $\theta$}} | D, M)=\frac{p(D | \mbox{{\boldmath $\theta$}}, M)p(\mbox{{\boldmath $\theta$}}|M)}{\int d\mbox{{\boldmath $\theta$}}p(D|\mbox{{\boldmath $\theta$}},M) p(\mbox{{\boldmath $\theta$}}|M)},\label{bayes}
\end{equation}
which gives the solution to the inverse problem in terms of the posterior probability distribution, $p(\mbox{{\boldmath $\theta$}} | D,M)$, that describes how probability is distributed among the possible values of the unknown parameter, {\boldmath $\theta$}, given the data $D$ and the assumed model $M$. The function $p(D |\mbox{{\boldmath $\theta$}}, M)$ is the likelihood of obtaining a data realization actually observed as a function of the parameter vector and provides a measure of how well the data are predicted by the model. The prior probability $p(\mbox{{\boldmath $\theta$}} |M)$ encodes any prior information we might have on the model parameters,  without taking into account the observed data. The denominator is the evidence,  an integral of the likelihood over the prior distribution. This quantity plays no role in parameter inference, but will become central in the model comparison described in Section~\ref{comparison}. 

\subsection{Longitudinally Stratified Loops}

For longitudinally stratified loops, model $M_1$, the forward problem is given in Eq.~(\ref{forward1}). To evaluate the likelihood, we assume the model is true. Then, the period ratio measurement ($r$) will differ from the prediction ($r_1$) because of measurement uncertainties ($e$), so that $r=r_1\pm e$. The probability of obtaining the measured value is equal to the probability of the error.  Assuming Gaussian errors, the likelihood for model $M_1$ is then expressed in the following manner

\begin{equation}\label{like1}
p(r | \eta,M_1)=\frac{1}{\sqrt{2\pi}\sigma}\exp \left[-\frac{(r-r_1)^2}{2\sigma^2}\right],
\end{equation}
with $\sigma^2$ the variance associated to the observed period ratio. In the following we assign observed period ratio errors to the standard deviation $\sigma$.

The prior indicates our level of knowledge (ignorance) before considering the observed data. We have adopted a uniform prior distribution for the unknown, $\eta$,  over a given range, so that we can write

\begin{equation}\label{prior1}
p(\eta|M_1)=\frac{1}{\eta^{\mathrm{max}}-\eta^{\mathrm{min}}} \mbox{\hspace{0.2cm}}\mbox{for}
\mbox{\hspace{0.2cm}}\eta^{\mathrm{min}}\leq \eta\leq \eta^{\mathrm{max}},
\end{equation}
and zero otherwise.  We only consider loops with density decreasing with height. We know from observations that very large coronal loops with the apex height above several atmospheric scale heights exist, so we consider the range $\eta\in[0,8]$ in the prior above.

Parameter inference is next performed using Bayes' theorem (\ref{bayes}). We use period ratio measurements reported in observations by \cite{verwichte04}. We must note that unambiguous identification of measured periods with particular overtones is a difficult task. Early observations were limited to the measurement of different periodicities at a single point, without spatial information \citep{verwichte04}. Even in the case in which information at different cuts along the loops is available \citep{demoortel07}, it is unclear how one should assign those periods to theoretical overtones. As our paper aims at providing a method to perform parameter inference and model comparison, we have considered period ratio values discussed in the review by \cite{andries09b}, noting that the reliability of the results  is closely related to that of the adopted assumptions and theoretical interpretations.

Figure~\ref{fig:posteriors}a shows posterior probability distributions for $\eta$ computed using Eq.~(\ref{bayes}) with likelihood and prior given by Eqs.~(\ref{like1}) and (\ref{prior1}), for two period ratio measurements by \cite{verwichte04}. Well constrained distributions are obtained.  For the measured period ratios  $r\sim0.79$  and $r\sim0.91$,  the inversion leads to density scale heights of $H=21$ Mm and $H=56$ Mm, respectively, for a loop with a height at the apex of  $L/\pi=70$ Mm. 

\subsection{Expanding Magnetic Loops}

For expanding loops, model $M_2$, the forward problem is given in Eq.~(\ref{forward2}). To evaluate the likelihood, we follow the same procedure as above, which leads to

\begin{equation}\label{like2}
p(r|\Gamma,M_2)=\frac{1}{\sqrt{2\pi}\sigma}\exp \left[-\frac{(r-r_2)^2}{2\sigma^2}\right].
\end{equation}

We also adopt a uniform prior distribution for the unknown $\Gamma$, so that we can write

\begin{equation}\label{prior2}
p(\Gamma|M_2)=\frac{1}{\Gamma^{\mathrm{max}}-\Gamma^{\mathrm{min}}} \mbox{\hspace{0.2cm}}\mbox{for}
\mbox{\hspace{0.2cm}}\Gamma^{\mathrm{min}}\leq\Gamma\leq\Gamma^{\mathrm{max}},
\end{equation}
and zero otherwise.  We consider $\Gamma\in[1,2.5]$.

Figure~\ref{fig:posteriors}b shows posterior probability distributions for $\Gamma$ computed using Eq.~(\ref{bayes}) with likelihood and prior given by Eqs.~(\ref{like2}) and (\ref{prior2}), for two period ratio measurements.  Again, well constrained distributions are obtained.  \cite{andries09b} discuss period ratio measurements in Table 1 by \cite{demoortel07}. Assuming that either the most power is in the fundamental mode or in the first overtone, mean values for the period ratio of  $r\sim1.07$  and $r\sim1.38$ are obtained. For those values, the inversion leads to tube expansion factors that are compatible with the estimates by \cite{klimchuk00} and \cite{watko00}.  Note however that, according to Fig~\ref{fig:posteriors}b, a period ratio of $r\sim 1.38$ requires an expansion of the tube by a factor of $\Gamma\sim 1.85$, while observations by \cite{watko00}  seem to indicate that in only very few cases does this parameter approach or exceed a value of $2$. 

For both theoretical models, the Bayesian framework makes use of all the available information in a consistent manner and enables us to consistently propagate errors from observations to inferred parameters.

\begin{figure*}[t]
   \centering
   \includegraphics[angle=0,scale=0.48]{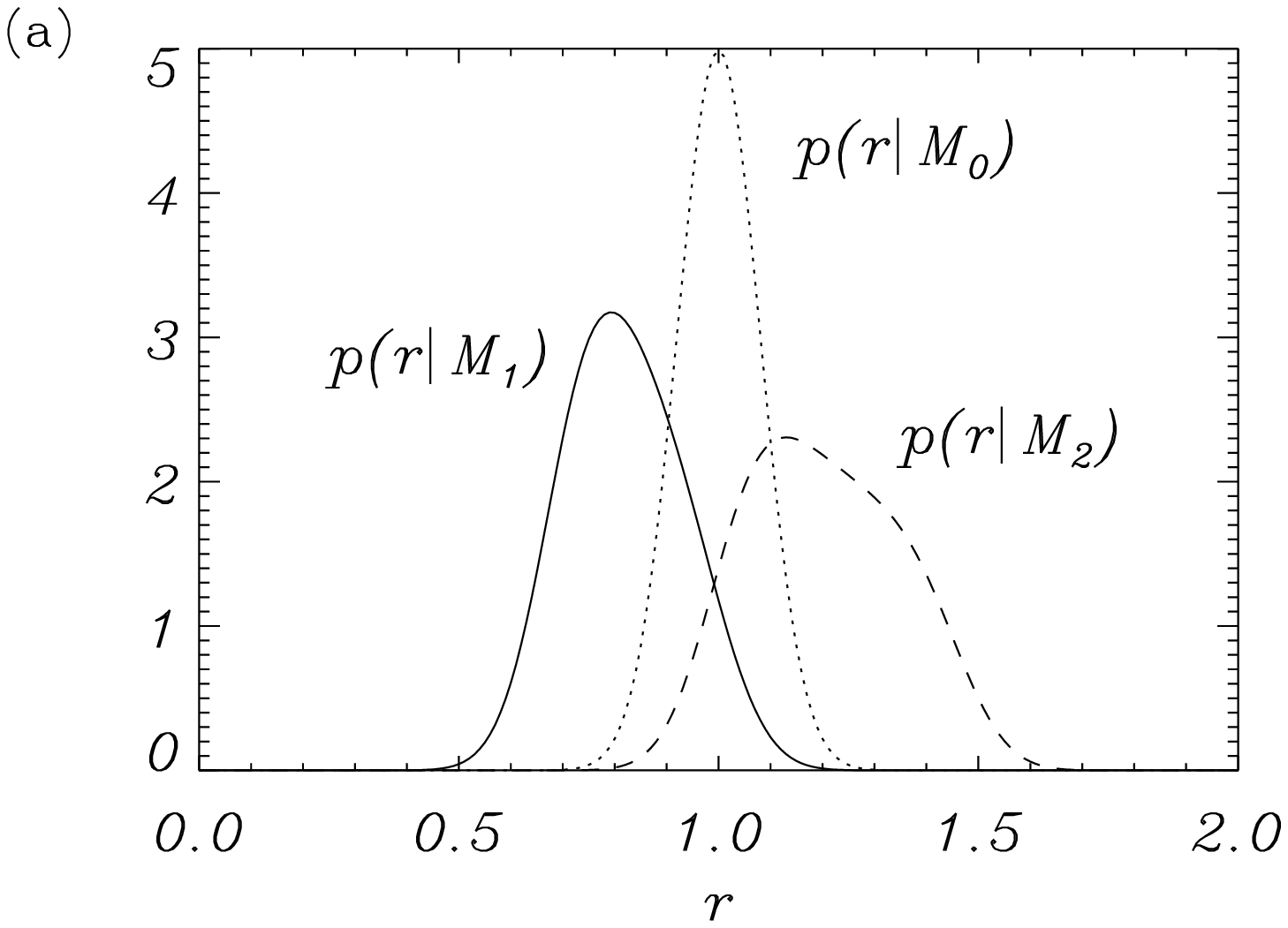} 
    \includegraphics[angle=0,scale=0.48]{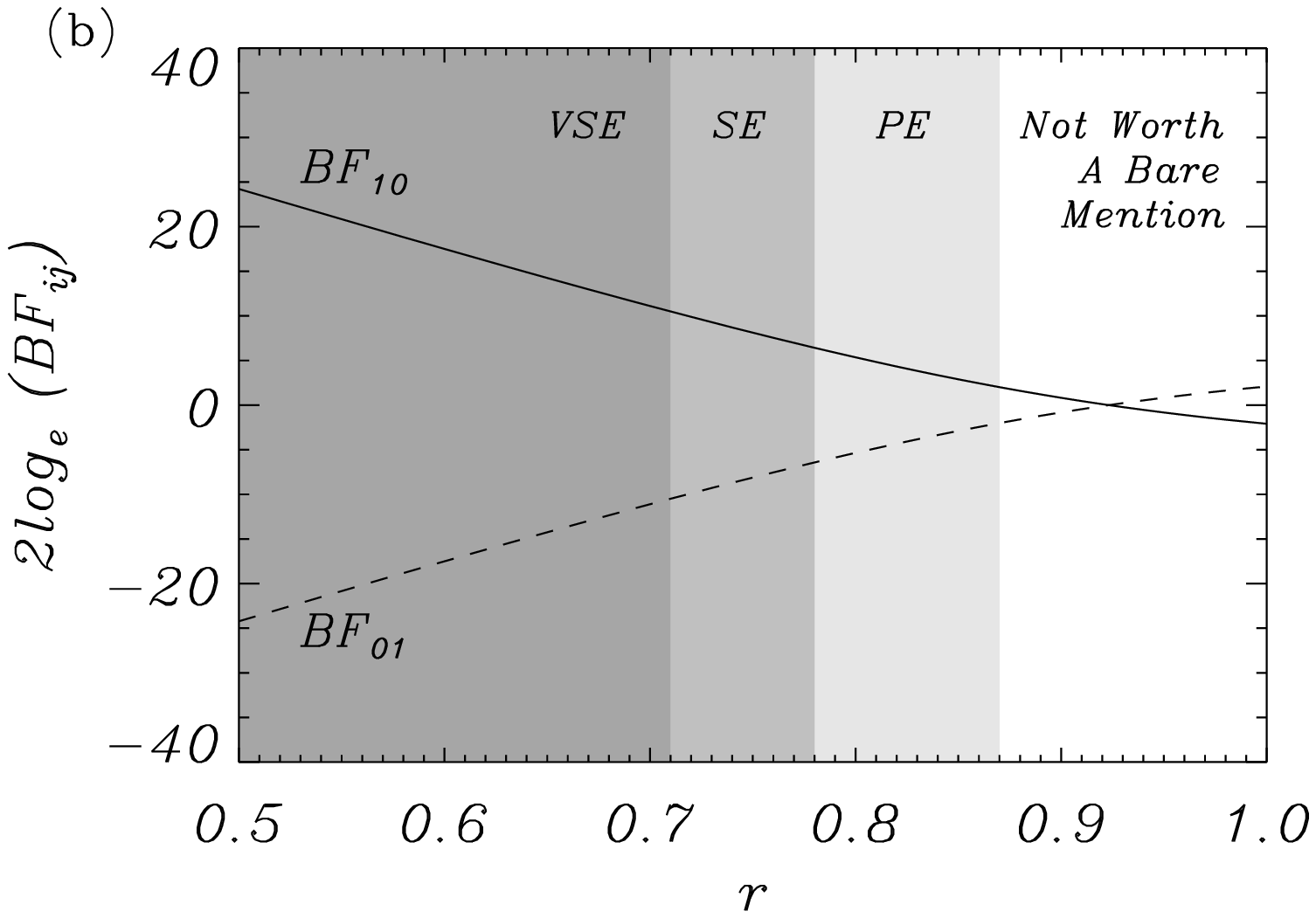}\\
    \includegraphics[angle=0,scale=0.48]{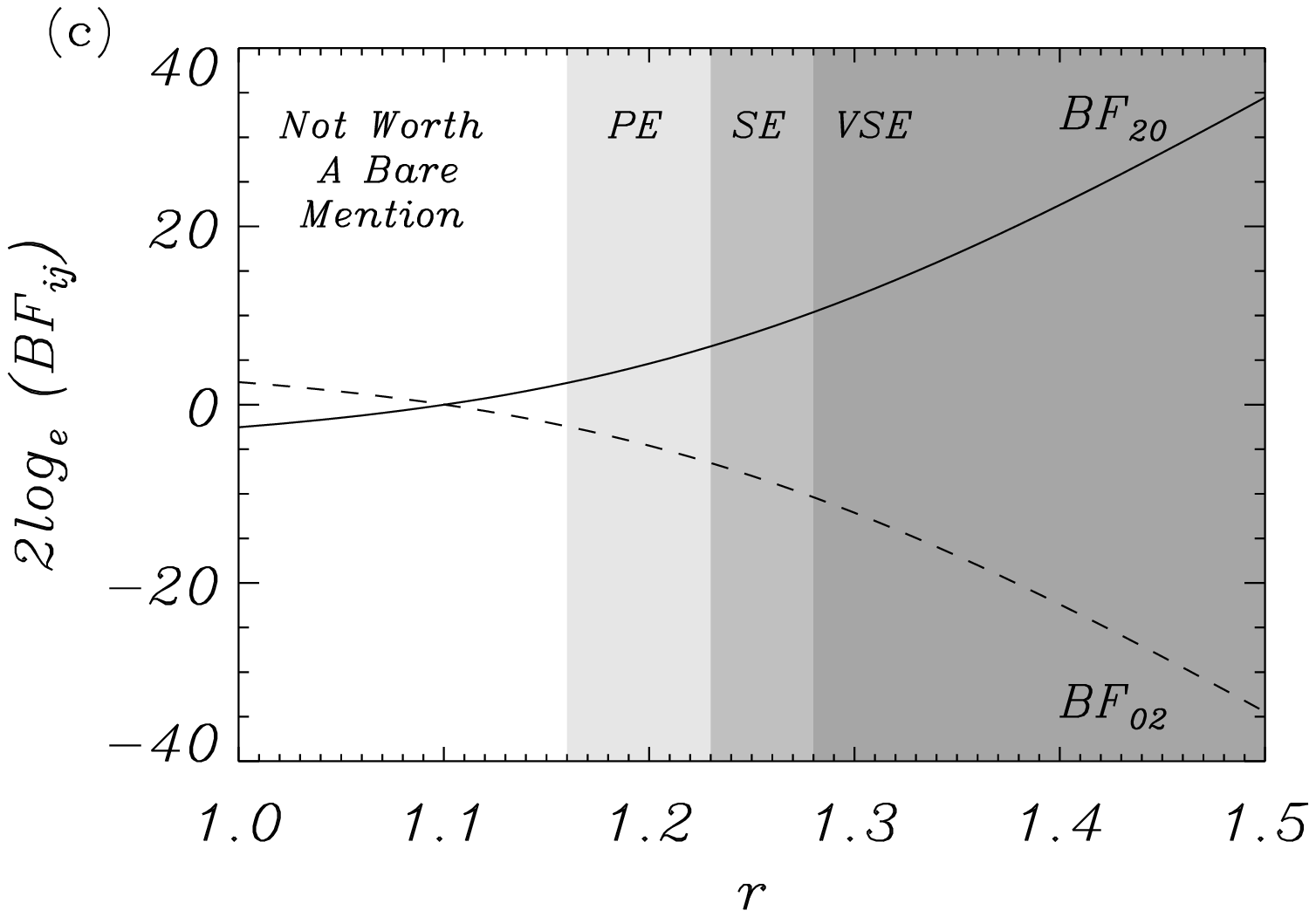} 
    \includegraphics[angle=0,scale=0.48]{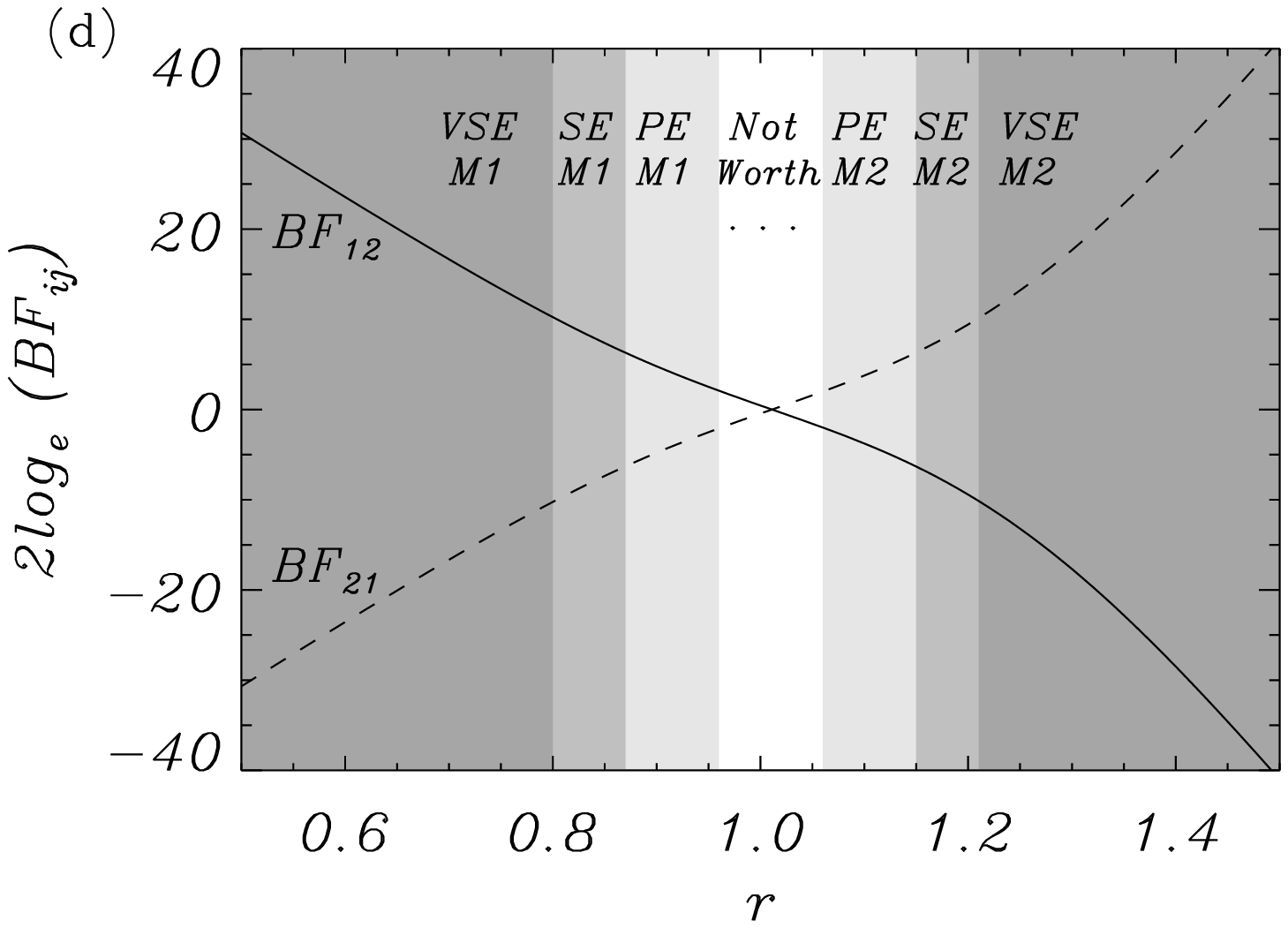}\\
    \caption{(a) Marginal likelihoods computed using Eq.~(\ref{marginal}) for models $M_0$, $M_1$ and $M_2$ as a function of the data. (b)-(d) Bayes factors computed using Eq.~(\ref{bf}) for the model comparisons between $M_1$ over $M_0$, $M_2$ over $M_0$, and $M_1$ over $M_2$.  In (b)-(d), white regions indicate evidence not worth more than a bare mention (2 log$_e BF\in[0,2]$); PE: positive evidence (2 log$_e BF\in[2,6]$); SE: strong evidence (2 log$_e BF\in[6,10]$);VSE: very strong evidence (2 log$_e BF >10$). Uniform priors in the ranges $\eta\in[0,8]$ and $\Gamma\in[1,2.5]$ have been taken. In all figures $\sigma=0.08$.}
   \label{fig:comparison1}
\end{figure*}

\section{Bayesian Model Comparison}\label{comparison}

Parameter inference was performed under the hypothesis that observed period ratios are the effect of either density stratification or magnetic tube expansion. The probability distributions in Figure~\ref{fig:posteriors} are conditional on the specific models used to explain the observations. The Bayesian framework  enables us to present different models to the same data and assess in a quantitative manner which one is favored by them. We have compared three models: $M_0$ for a uniform density and magnetic field strength tube, $M_1$ for a longitudinally stratified loop, and $M_2$ for a expanding magnetic loop.

As model $M_0$ predicts a period ratio of one, regardless of $\eta$ or $\Gamma$, we can write

\begin{equation}\label{like0}
p(r|M_0)=\frac{1}{\sqrt{2\pi}\sigma}\exp \left[-\frac{(r-1)^2}{2\sigma^2}\right],
\end{equation}
for the likelihood in this case. Note that $p(r|M_0)=p(r|\eta,M_0)=p(r|\Gamma,M_0)$.

To determine the plausibility of models $M_1$ and $M_2$ between them and with respect to model $M_0$, we  evaluate the posterior probabilities to ascertain the relative merits of two models, $M_i$ 
and $M_j$.  This is done by applying Bayes' theorem (Eq.~\ref{bayes}) to the two models and considering posterior ratios of the form \citep{jeffreys61}

\begin{equation}
\frac{p(M_i|r)}{p(M_j|r)}=\frac{p(r|M_i)}{p(r|M_j)}\frac{p(M_i)}{p(M_j)}.
\end{equation}
The first ratio on the right-hand side is the Bayes factor. It expresses how well the observed data are predicted by model $M_i$, compared to model $M_j$. The second ratio, the prior odds ratio, measures how much our initial beliefs favored $M_i$ over $M_j$, before considering the data.
As we have no particular a priori preference for one model over the other, before considering the data, we take $p(M_i)=p(M_j)=1/2$. Our assessment of the plausibility of models is then based on the computation of the Bayes factor of $M_i$ against $M_j$ given by

\begin{equation}\label{bf}
BF_{ij}=\frac{p(r|M_i)}{p(r|M_j)}.
\end{equation}

In model selection, we are interested in the most probable model, independently of the parameters, i.e., we should marginalise out all parameters. This is achieved by performing an integral of the likelihood over the full parameter space.  The marginal likelihood for a given model $M_i$ is then
\begin{equation}\label{marginal}
p(r|M_i)=\int^{\theta^{\mathrm max}}_{\theta^{\mathrm min}}p(r,\theta|M_i)d\theta=\int^{\theta^{\mathrm max}}_{\theta^{\mathrm min}}p(r|\theta,M_i)p(\theta|M_i)d\theta,
\end{equation}
where $\theta\in[\theta^{\mathrm min},\theta^{\mathrm max}]$ represents the particular parameter of the model and we have used the product rule to expand the probability of $r$ and $\theta$, given model $M_i$. 

Figure~\ref{fig:comparison1}a displays the marginal likelihoods for the three considered models. For a given observed period ratio, the plausibility of one model over the other is given by the ratio of these two quantities at the measured period ratio $r$. The uniform model has a marginal likelihood that is maximum at one. The model with density stratification is clearly favored for observations of period ratios below unity. Figure~\ref{fig:comparison1}b quantifies the relative performance of models $M_1$ and $M_0$ by computing the Bayes factor. The lower $r$, the more plausible $M_1$ is against $M_0$. Jeffreys' scale \citep{jeffreys61,kass95} assigns different levels of evidence to the values of the Bayes factor in natural logarithm units.  They are shown using different darkening options in Figures~\ref{fig:comparison1}b-d. According to  Figure~\ref{fig:comparison1}b, and given the assumed uncertainty of $\sigma=0.08$, a period ratio measurement should be considered as positive evidence for model $M_1$ against model $M_0$ only if it is  below 0.87. As strong evidence, only if it is below 0.78.  A period ratio below 0.71 indicates very strong evidence for model $M_1$. 

A similar comparison was made for models $M_2$ and $M_0$ (Figure~\ref{fig:comparison1}c). The model for magnetic tube expansion is clearly favored for observations of period ratios above unity. Model $M_0$ cannot be ruled out by the simple fact that $r>1$. Positive evidence for model $M_2$ against model $M_0$ exists for period ratio measurements above 1.16. If $r>1.23$, there is  strong evidence for  model $M_2$. A period ratio above 1.28 indicates very strong evidence for model $M_2$. 

Finally, we have compared model $M_1$ against model $M_2$, in view of observed data. Figure~\ref{fig:comparison1}d shows that the evidence is the same for both models if the period ratio is unity. Around this value, there is a region, $0.96<r<1.06$, in which no firm conclusion can be established. Positive, strong, and very strong evidence for $M_1$ occur below 0.96, 0.87, and 0.80, respectively. Positive, strong, and very strong evidence for $M_2$ occur above 1.06, 1.15, and 1.21, respectively.

In our model comparison, $\sigma=0.08$ has been selected so as to clearly show the different regimes for the evidence. An increase (decrease) of $\sigma$ produces a decrease (increase) of any evidence.
The inference in Figure~\ref{fig:posteriors}a for $r=0.91$ with uncertainty of $\sigma=0.04$ falls into the region of positive evidence for hypothesis $M_1$. The inference in Figure~\ref{fig:posteriors}b for $r=1.07$ should have an uncertainty of $\sigma=0.03$ (close to the reported error) to be considered done under positive evidence for hypothesis $M_2$.

\section{Conclusions}
We have presented a simple and straightforward technique to perform parameter inference and model comparison using period ratios of kink oscillations in the Bayesian framework. Parameter inference enable us to obtain estimates for the coronal density scale height and the magnetic loop expansion, using all the available information and with correctly propagated uncertainties. Density scale height estimates are compatible with previous studies. Magnetic tube expansion factors are compatible with the estimates by \cite{klimchuk00} and \cite{watko00}. Even if Eqs.~(\ref{forward1}) and (\ref{forward2}) imply well-posed
inversion problems from a mathematical point of view, the unknown parameters cannot be obtained uniquely, due to the uncertainties associated to observed data. The Bayesian framework enables us to consistently deal with this problem.

Bayesian model comparison enable us to assess the plausibility of the inferences, which are conditional on theoretical models. Deviations of the period ratio below (above) unity do not necessarily imply density stratification (tube expansion) to be preferred in front of the null hypothesis.
The degree of evidence for the two theories so far invoked can be quantitatively assessed with the use of Bayes factors, to precisely decide which one of the two hypotheses is more plausible, for a given  measured period ratio and the associated uncertainty.

Our analysis provides a simple and easy to use method to perform inference and model comparison in the presence of incomplete and uncertain information.  Measured period ratios and their uncertainties determine the strength of  the evidence in favour of a particular hypothesis and, therefore, the support of the performed inferences. 

The Bayesian formalism for inference and model comparison is the only fully correct way we have to get information about physical parameters and the plausibility of hypotheses from observations \citep[see e.g.][]{trotta08,vontoussaint11} under incomplete and uncertain information. \cite{arregui11b} performed the first Bayesian parameter inference in coronal seismology. This paper presents the first application of Bayesian techniques to model comparison. When analytical forward problems are available, both parameter inference and model comparison reduce to a simple and straightforward evaluation of the marginal posteriors and the Bayes factors, that can be obtained by solving simple integrals. The methods outlined in this paper can be directly applied to most of the seismology inversion problems in which other physical effects and parameters are involved. For instance, additional effects that influence period ratios could be compared to the already considered hypotheses.

\acknowledgments
I.A. and A.A.R. acknowledge support by Ram\'on y Cajal Fellowships by the Spanish Ministry of Economy and Competitiveness (MINECO). I.A. acknowledges the support from the Spanish MICINN/MINECO and FEDER funds through project AYA2011-22846. A.A.R. and A.J.D. acknowledge support by the Spanish MINECO through project AYA2010-18029 (Solar Magnetism and Astrophysical Spectropolarimetry). A.A.R. also acknowledges support from the Consolider-Ingenio 2010 CSD2009-00038 project. We are grateful to the referee and to Ram\'on Oliver for valuable comments.

\end{document}